\documentstyle[aps,pra]{revtex}
\begin{document}
\title{Conditions for optimal construction of two-qubit non-local gates}
\author{Yong-Sheng Zhang\thanks{%
Electronic address: yshzhang@ustc.edu.cn}, Ming-Yong Ye, and Guang-Can Guo%
\thanks{%
Electronic address: gcguo@ustc.edu.cn}}
\address{Laboratory of Quantum Information, University of Science and Technology of\\
China, Hefei 230026, People's Republic of China\vspace{0.5in}}
\maketitle

\begin{abstract}
\baselineskip12pt Optimal implementation of quantum gates is crucial for
designing a quantum computer. The necessary condition for optimal
construction of a two-qubit unitary operation is obtained. It can be proved
that the B gate is the unique gate that can construct a two-qubit universal
circuit with only two applications, i.e. this condition is also sufficient
in the case of two applications of the elementary two-qubit gate. It is also
shown that one half of perfect entanglers can not simulate an arbitrary
two-qubit gate with only 3 applications.

PACS number(s): 03.67.Lx, 03.67.-a \medskip
\end{abstract}

\baselineskip12ptQuantum computation can be described by unitary matrices.
In order to effect a quantum computation on a quantum computer, one must
decompose the corresponding unitary matrix into a quantum circuit which
consists of elementary quantum gates\cite{Gate}. It has been shown that any
interaction that can create entanglement between any pair of qudits (qubits)
is universal for quantum computation together with one-qudit (qubit) gates%
\cite{gate2,Bennett,Vidal1}. Many efforts are devoted to the optimal
construction of explicit quantum circuits for arbitrary unitary operation.
The current standard paradigm is based on a combination of quantum
Controlled-NOT (CNOT) gates between pairs of qubits and single-qubit gates%
\cite{Barenco}. Recent work shows that the CNOT gate is also one of the most
efficient quantum gates known, in that just three applications supplemented
with local gates can implement any arbitrary two-qubit operation\cite{Vidal}.

However, a practical Hamiltonian can not necessarily be efficient to
construct the CNOT gate. It is possible to construct different elementary
gates that depend on the given Hamiltonian. Questions are raised that are
there any gates which have more efficiency than CNOT, or what is the optimal
construction of a circuit if the elementary two-qubit gate is not a standard
CNOT gate? J. Zhang {\it et al}. have proposed optimal construction with B
gate and Controlled-Unitary (Controlled-U) gate respectively\cite
{Zhang1,Zhang2}. Bremner {\it et al.} have given an operable construction
with an arbitrary two-qubit elementary gate, though this construction is not
always optimal\cite{Bremner}.

In this paper, the necessary condition of optimal construction for a
two-qubit gate with a general elementary entangling gate is proposed.
Nontrivial results in the special cases of two or three applications of the
elementary gate are also proposed.

It is known that an arbitrary two-qubit unitary transformation $U$ can be
decomposed into this form\cite{decompose,Kraus}. 
\begin{eqnarray}
U &=&\left( A_1\otimes B_1\right) e^{iH_U}\left( A_2\otimes B_2\right) , 
\eqnum{1} \\
H_U &=&c_1\sigma _x\otimes \sigma _x+c_2\sigma _y\otimes \sigma _y+c_3\sigma
_z\otimes \sigma _z.  \nonumber
\end{eqnarray}
where $\sigma _x,\sigma _y,\sigma _z$ are the Pauli matrices, $A_j,B_j$ are
single qubit gates, and $\frac \pi 4\geq c_1\geq c_2\geq \left| c_3\right| $%
. The non-local content $\phi \left( U\right) =\lambda \left( H_U\right) $
are 
\begin{eqnarray}
\lambda _1 &=&c_1+c_2-c_3,  \eqnum{2} \\
\lambda _2 &=&c_1-c_2+c_3,  \nonumber \\
\lambda _3 &=&-c_1+c_2+c_3,  \nonumber \\
\lambda _4 &=&-c_1-c_2-c_3,  \nonumber
\end{eqnarray}
where $\lambda \left( A\right) $ denotes the vector whose entries are
eigenvalues of Hermitian matrix $A,$ arranged into non-increasing order. In
the following, we will denote $e^{iH_U}$ as $U_d\left( c_1,c_2,c_3\right) $.

Our first result is based on the lemmas which were proposed by Childs {\it %
et al. }\cite{Childs} and the following theorem of K. Fan \cite{KyFan}.

{\it Lemma 1}. Let $H$, $K$ be Hermitian matrices. Then 
\begin{equation}
\lambda \left( H+K\right) \prec \lambda \left( H\right) +\lambda \left(
K\right) ,  \eqnum{3}
\end{equation}
where the majorization relation whose notation is $\prec $ will be described
as follows. Suppose $x=\left( x_1,\cdots ,x_D\right) $ and $y=\left(
y_1,\cdots ,y_D\right) $ are two D-dimensional real vectors. $x$ is
majorized by $y$, written $x\prec y$, if 
\begin{equation}
\sum_{j=1}^kx_j^{\downarrow }\leq \sum_{j=1}^ky_j^{\downarrow }  \eqnum{4}
\end{equation}
for $k=1,\cdots ,D-1$, and the inequality holds with equality where $k=D$.
Here $\downarrow $ denotes the components of a vector rearranged into
non-increasing order.

Combining Lemma 4, Lemma 6 of Ref. \cite{Childs}, and the above Lemma 1, it
is not difficult to obtained the following theorem.

{\it Theorem 2}. Let $U_1,U_2$ be two-qubit unitary operations, then 
\begin{equation}
\phi \left( U_1U_2\right) \prec \phi \left( U_1\right) +\phi \left(
U_2\right) .  \eqnum{5}
\end{equation}

{\it Proof}. We denote $U_0=U_1U_2.$ According to Lemma 4 of Ref. \cite
{Childs}, there exist $H_j$'s $\left( j=0,1,2\right) $ that satisfy 
\[
U_j\widetilde{U}_j=e^{2iH_j}\text{ and }\lambda \left( H_j\right) =\phi
\left( U_j\right) . 
\]
According to Lemma 6 of Ref. \cite{Childs}, there exist $K_1,$ and $K_2$
that satisfy 
\[
\lambda \left( U_0\widetilde{U}_0\right) =\lambda \left( e^{2i\left(
K_1+K_2\right) }\right) , 
\]
where $\lambda \left( K_1\right) =\lambda \left( H_1\right) $ and $\lambda
\left( K_2\right) =\lambda \left( H_2\right) .$ According to Lemma 4 of Ref. 
\cite{Childs}, there must be 
\[
\lambda \left( K_1+K_2\right) =\phi \left( U_0\right) +\pi \overrightarrow{m}%
, 
\]
where $\overrightarrow{m}$ is an integer vector. With the above K. Fan's
Lemma, it can be obtained that 
\begin{eqnarray}
\phi \left( U_1\right) +\phi \left( U_2\right) &=&\lambda \left( K_1\right)
+\lambda \left( K_2\right)  \eqnum{6} \\
&\succ &\lambda \left( K_1+K_2\right)  \nonumber \\
&=&\phi \left( U_0\right) +\pi \overrightarrow{m}.  \nonumber
\end{eqnarray}
To satisfy this relation, there must be (Here $\overrightarrow{m}$ is
written in non-increasing order but $\phi \left( U_0\right) $ is not
necessarily written in this order.) 
\[
\overrightarrow{m}=\left( 0,0,0,0\right) \text{ or }\overrightarrow{m}%
=\left( 1,0,0,-1\right) . 
\]
It is not difficult to obtained that 
\[
\phi ^{\prime }\left( U_0\right) +\pi \left( 1,0,0,-1\right) \succ \phi
\left( U_0\right) , 
\]
where $\phi ^{\prime }\left( U_0\right) $ is any permutation of $\phi \left(
U_0\right) .$ So 
\[
\phi \left( U_1\right) +\phi \left( U_2\right) \succ \phi \left(
U_1U_2\right) . 
\]

Though we can get conditions for gate simulation directly in the same way of
Hamiltonian simulation \cite{Childs} with theorem 2, the next lemma gives
more strict constraints than Hamiltonian simulation.

{\it Lemma 3}. Two consecutive Swap-class gates are equivalent to local
unitary operation.

{\it Proof}. It is known any Swap-class gate is equivalent to $U_d\left( 
\frac \pi 4,\frac \pi 4,\frac \pi 4\right) $. So 
\begin{eqnarray*}
&&\left( A_1\otimes B_1\right) U_d\left( \frac \pi 4,\frac \pi 4,\frac \pi 4%
\right) \left( U_A\otimes U_B\right) U_d\left( \frac \pi 4,\frac \pi 4,\frac %
\pi 4\right) \left( A_2\otimes B_2\right) \\
&=&\left( A_1^{\prime }\otimes B_1^{\prime }\right) \left( Swap\right)
\left( U_A^{\prime }\otimes U_B^{\prime }\right) \left( Swap\right) \left(
A_2^{\prime }\otimes B_2^{\prime }\right) \\
&=&\left( A_1^{\prime }U_B^{\prime }A_2^{\prime }\right) \otimes \left(
A_2^{\prime }U_A^{\prime }A_1^{\prime }\right) .
\end{eqnarray*}

{\it Theorem 4}. The necessary conditions that $n$ applications of a $%
U_d\left( c_1,c_2,c_3\right) -$class gate can simulate any two-qubit unitary
operations are 
\begin{eqnarray}
n\left( c_1+c_2-\left| c_3\right| \right) &\geq &\frac{3\pi }4,  \eqnum{7} \\
n\left( c_1-c_2-\left| c_3\right| +\frac \pi 4\right) &\geq &\frac{3\pi }4. 
\nonumber
\end{eqnarray}

{\it Proof}. Since we have supposed that $n$ applications of the $U_d\left(
c_1,c_2,c_3\right) -$class gate can simulate any two-qubit gate, for
continuity, it should be able to simulate the vertex points of the geometric
representation of non-local two-qubit unitary operation, i.e. it should be
able to simulate the vertex point $U_d\left( \frac \pi 4,\frac \pi 4,\frac %
\pi 4\right) $ and $U_d\left( \frac \pi 4,\frac \pi 4,-\frac \pi 4\right) $%
(Though they are the same gate, but their neighbor points in the tetrahedral
representation are different from each other.). So we can obtain 
\begin{eqnarray}
n\phi \left( U_d\left( c_1,c_2,c_3\right) \right) &\succ &\phi \left(
U_d\left( \frac \pi 4,\frac \pi 4,\frac \pi 4\right) \right) ,  \eqnum{8} \\
n\phi \left( U_d\left( c_1,c_2,c_3\right) \right) &\succ &\phi \left(
U_d\left( \frac \pi 4,\frac \pi 4,-\frac \pi 4\right) \right) .  \nonumber
\end{eqnarray}
There must be 
\[
n\left( c_1+c_2-\left| c_3\right| \right) \geq \frac{3\pi }4. 
\]

On the other hand, using lemma 3, we can get that if $n$ applications of $%
U_d\left( c_1,c_2,c_3\right) $ can simulate any two-qubit gate, 
\[
\overline{U_d}\left( c_1,c_2,c_3\right) =U_d\left( \frac \pi 4-\left|
c_3\right| ,\frac \pi 4-c_2,sign\left( c_3\right) \left( c_1-\frac \pi 4%
\right) \right) 
\]
can also construct any two-qubit unitary operation with $n$ applications.
This is because that $U_d\left( \frac \pi 4-\left| c_3\right| ,\frac \pi 4%
-c_2,sign\left( c_3\right) \left( c_1-\frac \pi 4\right) \right) $ is
equivalent to $U_d\left( c_1,c_2,c_3\right) U_d\left( \frac \pi 4,\frac \pi 4%
,-sign\left( c_3\right) \frac \pi 4\right) $, and two consecutive $U_d\left( 
\frac \pi 4,\frac \pi 4,-sign\left( c_3\right) \frac \pi 4\right) $ are
equivalent to local unitary operations (If $n$ is an odd number, we can
multiply a SWAP-class gate on the gate to be simulated. It does not affect
our proof since the gate to be simulated is an arbitrary two-qubit gate).
Note here sign function is defined as 
\begin{eqnarray}
sign\left( x\right) &=&1,\text{when }x\geq 0,  \eqnum{9} \\
sign\left( x\right) &=&-1,\text{when }x<0.  \nonumber
\end{eqnarray}
From this discussion, it can be obtained that 
\begin{eqnarray}
n\phi \left( \overline{U_d}\left( c_1,c_2,c_3\right) \right) &\succ &\phi
\left( U_d\left( \frac \pi 4,\frac \pi 4,\frac \pi 4\right) \right) , 
\eqnum{10} \\
n\phi \left( \overline{U_d}\left( c_1,c_2,c_3\right) \right) &\succ &\phi
\left( U_d\left( \frac \pi 4,\frac \pi 4,-\frac \pi 4\right) \right) . 
\nonumber
\end{eqnarray}
So we can get that 
\[
n\left( c_1-c_2-\left| c_3\right| +\frac \pi 4\right) \geq \frac{3\pi }4. 
\]

In the situation that $c_2=c_3=0,$ i.e. the elementary gate is a
controlled-U gate, it can be obtained that the minimum applications required
to implement any arbitrary two-qubit gate together with local gates is $%
\left\lceil \frac{3\pi }{4c_1}\right\rceil $ ($\left\lceil x\right\rceil $
is the minimum integer number that is not smaller than $x.$). It is just the
same condition proposed by J. Zhang {\it et al.}\cite{Zhang2}, and it was
also proved to be sufficient there.

{\it Theorem} 5. B gate is the {\it unique} gate (up to local unitary
operations) that can simulate any two-qubit gate with only two applications.

{\it Proof}. Here the B gate is just the $U_d\left( \frac \pi 4,\frac \pi 8%
,0\right) $ gate which was proposed by J. Zhang {\it et al}.\cite{Zhang1}.
According to Theorem 4, if two applications of $U_d\left( c_1,c_2,c_3\right) 
$ can simulate any two-qubit gate, three parameters of the gate must satisfy 
\begin{eqnarray}
c_1+c_2-\left| c_3\right| &\geq &\frac{3\pi }8,  \eqnum{11} \\
c_1-c_2-\left| c_3\right| &\geq &\frac \pi 8.  \nonumber
\end{eqnarray}
Since $\frac \pi 4\geq c_1\geq c_2\geq \left| c_3\right| $, there must be 
\begin{equation}
c_1=\frac \pi 4,c_2=\frac \pi 8,c_3=0.  \eqnum{12}
\end{equation}

We can understand that $c_3$ must be zero in another point of view. For
continuity, two applications of this gate should be able to construct the
identity gate, i.e. $U_d\left( c_1,c_2,c_3\right) $ should be equivalent to $%
U_d^{-1}\left( c_1,c_2,c_3\right) $. Comparing the Makhlin's invariants\cite
{Mak,Zhang3} of $U_d\left( c_1,c_2,c_3\right) $ and $U_d^{-1}\left(
c_1,c_2,c_3\right) $, we can obtain that $c_3=0$. The sufficiency of this
theorem has been proved by J. Zhang {\it et al}. \cite{Zhang1}.

Applying theorem 4 in the case of $n=3$, we can obtain the following
corollary directly.

{\it Corollary 6}. The necessary conditions that a gate $U_d\left(
c_1,c_2,c_3\right) $ can simulate any arbitrary two-qubit unitary operation
with three applications are 
\begin{eqnarray}
c_1+c_2-\left| c_3\right| &\geq &\frac \pi 4,  \eqnum{13} \\
c_1-c_2-\left| c_3\right| &\geq &0.  \nonumber
\end{eqnarray}

As mentioned above, the CNOT, Double CNOT (DCNOT) and Super controlled gates%
\cite{Ye} all can construct any two-qubit gate with three applications. They
all satisfy this condition. However, one half of perfect entanglers \cite
{Kraus} which can generate maximal entanglement states from product states
can not construct an arbitrary two-qubit gate with only 3 applications. This
result was shown in Fig. 1. (For convenience, we only depict the tetrahedron 
$OACF$: $\frac \pi 4\geq c_1\geq c_2\geq c_3\geq 0.$ It is not difficult to
depict the tetrahedron $\frac \pi 4\geq c_1\geq c_2\geq -c_3\geq 0$ by
symmetry.). It is somewhat a surprising result, since that all perfect
entanglers have the same entangling ability as CNOT without auxiliary
system. For example, $\sqrt{SWAP}$ gate which is represented by $U_d\left( 
\frac \pi 8,\frac \pi 8,\frac \pi 8\right) $ does not satisfy the condition
but it is a perfect entangler. $\sqrt{SWAP}$ can be produced by exchange
interaction Hamiltonian $J\overrightarrow{S_1}\cdot \overrightarrow{S_2},$
which is typical in solid systems. In addition, it is the most powerful gate
in this class, but at least 6 applications of it is required to implement an
arbitrary two-qubit unitary operation.

\begin{center}
{\bf Figure 1.}
\end{center}

We conjecture that 
\begin{equation}
P=\min \left\{ c_1+c_2-\left| c_3\right| ,\frac \pi 4+c_1-c_2-\left|
c_3\right| \right\}  \eqnum{14}
\end{equation}
can measure the ability of a gate to construct a universal two-qubit unitary
operation. For this measurement, the B gate is the most powerful one, the
SWAP gate and local unitary gate have the smallest ability. In general, at
least $\left\lceil \frac{3\pi }{4P}\right\rceil $ applications is needed for
a $U_d\left( c_1,c_2,c_3\right) $-class gate to construct an arbitrary
two-qubit gate.

An arbitrary two-qubit gate can also be constructed with different types of
non-local gates. For example, three parameter-tunable $\left( Swap\right)
^\alpha $ gate can simulate an arbitrary two-qubit gate \cite{Fan} but three
parameter-fixed $\left( Swap\right) ^\alpha $ gate can not. Hence, it was
also proved that 3 applications of a nontrivial Hamiltonian can simulate any
two-qubit unitary operations \cite{Zhang3}. However, it is more difficult to
solve the optimality for different fixed gates acting as elementary gates in
general. Though we can introduce the necessary conditions by the same way as
we have mentioned above, it is not easy to get an exact form. But in some
simple cases such as controlled-U gate, we can get the following result.

{\it Theorem 7}. \cite{Ye} If a two-qubit gate $U_d\left( c_1,c_2,0\right) $
can be simulated by $U_d\left( \gamma _1,0,0\right) $ and $U_d\left( \gamma
_2,0,0\right) $, the parameters must satisfy the following conditions. 
\begin{eqnarray}
\gamma _1+\gamma _2 &\geq &c_1+c_2,  \eqnum{15} \\
\left| \gamma _1-\gamma _2\right| &\leq &c_1-c_2.  \nonumber
\end{eqnarray}

{\it Proof}: Since that 
\begin{equation}
\phi \left( U_d\left( \gamma _1,0,0\right) U_d\left( \gamma _2,0,0\right)
\right) \succ \phi \left( U_d\left( c_1,c_2,0\right) \right) ,  \eqnum{16}
\end{equation}
it can be obtained that $\gamma _1+\gamma _2\geq c_1+c_2.$

On the other hand, if $U_d\left( \gamma _1,0,0\right) $ and $U_d\left(
\gamma _2,0,0\right) $ can simulate $U_d\left( c_1,c_2,0\right) $, $%
U_d\left( c_1,c_2,0\right) $ and $U_d^{-1}\left( \gamma _2,0,0\right) $
which is equivalent to $U_d\left( \gamma _2,0,0\right) $ also can simulate $%
U_d\left( \gamma _1,0,0\right) $. Using the lemma 3, it can be obtained that 
\begin{equation}
\phi \left( U_d\left( c_1,c_2,0\right) \overline{U_d}\left( \gamma
_2,0,0\right) \right) \succ \phi \left( U_d\left( \gamma _1,0,0\right)
\right) ,  \eqnum{17}
\end{equation}
that is $c_1-c_2\geq \gamma _1-\gamma _2.$ Similarly, $c_1-c_2\geq \gamma
_2-\gamma _1$. So there should be $\left| \gamma _1-\gamma _2\right| \leq
c_1-c_2.$

Note J. Zhang {\it et al}. \cite{Zhang2} have proved that two controlled-U
gate can not construct a $U_d\left( c_1,c_2,c_3\right) $ gate with $c_3\neq
0 $.

It is known that realizing quantum computation needs to decompose the
desired unitary operation into elementary gates. Earlier work to complete
this decomposition is to find the universal two-qubit gate, and this needs
infinite time use of a fixed two-qubit elementary gate in almost all of the
case. Another more practical way is to use a fixed non-trivial two-qubit
gate (excluding SWAP and identity gates) together with arbitrary one-qubit
gates. This method can construct an arbitrary unitary operation with finite
time using of elementary gates. On the other hand, the realization of some
``standard'' gate, e.g. CNOT, is not always efficient by different
Hamiltonian. Investigation of the optimal construction with general
two-qubit entangling gates is required. The necessary condition for optimal
construction of a two-qubit unitary operation is proposed in this paper. It
can be proved that this condition is also sufficient in the case of two
applications of the elementary two-qubit gate, i.e. the B gate is the unique
gate that can construct a two-qubit universal circuit with only two
applications. It is also shown that one half of perfect entanglers can not
simulate any two-qubit gate with only 3 applications. However, there are
many open questions of the optimal simulation, e.g. the sufficient
conditions of Theorem 4, i.e. the optimal operable construction is desired.

Simulating a gate by another gate is also different from Hamiltonian
simulation which allows infinitely many steps of evolution \cite
{Bennett,Vidal1,Hami,Zeier}. The gate simulation, i.e. finite times of
switch on the Hamiltonian is more practical in realization of an actual
quantum process.

This work was supported by the National Fundamental Research Program
(2001CB309300), the National Natural Science Foundation of China (No.
10304017), the Innovation Funds from Chinese Academy of Sciences
(CAS).


\medskip

\begin{references}
\bibitem{Gate}  \baselineskip12ptD. Deutsch {\it et al}., Proc. R. Soc.
Lond. A {\bf 449}, 669 (1995); D. P. DiVincenzo, Phys. Rev. A {\bf 51}, 1015
(1995); T. Sleator and H. Weinfurter, Phys. Rev. Lett. {\bf 74}, 4087
(1995); A. Barenco, Proc. R. Soc. Lond. A {\bf 449}, 678 (1995); S. Lloyd,
Phys. Rev. Lett. {\bf 75}, 346 (1995).

\bibitem{gate2}  J. L. Dodd, M. A. Nielsen, M. J. Bremner, and R. T. Thew,
Phys. Rev. A {\bf 65}, 040301 (2002); W. D\"{u}r, G. Vidal, J. I. Cirac, N.
Linden, and S. Popescu, Phys. Rev. Lett. {\bf 87}, 137901 (2001); M. A.
Nielsen, M. J. Bremner, J. L. Dodd, A. M. Childs, and C. M. Dawson, Phys.
Rev. A {\bf 66}, 062317 (2002); P. Wocjan, D. Janzing, and T. Beth, Quantum
Inf. Comput. {\bf 2}, 117 (2002).

\bibitem{Bennett}  C. H. Bennett {\it et al}., Phys. Rev. A {\bf 66}, 012305
(2002).

\bibitem{Vidal1}  G. Vidal and J. I. Cirac, Phys. Rev. A {\bf 66}, 022315
(2002).

\bibitem{Barenco}  A. Barenco {\it et al}., Phys. Rev. A {\bf 52}, 3457
(1995).

\bibitem{Vidal}  G. Vidal and C. M. Dawson, Phys. Rev. A {\bf 69}, 010301
(2004); F. Vatan and C. Williams, Phys. Rev. A {\bf 69}, 032315 (2004); V.
V. Shende, I. L. Markov, and S. S. Bullock, Phys. Rev. A {\bf 69}, 062321
(2004).

\bibitem{Zhang1}  J. Zhang, J. Vala, S. Sastry, and K. B. Whaley, Phys. Rev.
Lett. {\bf 93}, 020502 (2004).

\bibitem{Zhang2}  J. Zhang, J. Vala, S. Sastry, and K. B. Whaley, Phys. Rev.
A {\bf 69}, 042309 (2004).

\bibitem{Bremner}  M. J. Bremner {\it et al}., Phys. Rev. Lett. {\bf 89},
247902 (2002).

\bibitem{decompose}  N. Khaneja, R. Brockett, and S. J. Glaser, Phys. Rev. A 
{\bf 63}, 032308 (2001).

\bibitem{Kraus}  B. Kraus and J. I. Cirac, Phys. Rev. A {\bf 63}, 062309
(2001).

\bibitem{KyFan}  K. Fan, Proc. Natl. Acad. Sci. U.S.A. {\bf 35}, 131 (1949).

\bibitem{Childs}  A. M. Childs, H. L. Haselgrove, and M. A. Nielsen, Phys.
Rev. A {\bf 68}, 052311 (2003).

\bibitem{Mak}  Y. Makhlin, Quant. Inf. Proc. {\bf 1}, 243 (2002);
quant-ph/0002045.

\bibitem{Zhang3}  J. Zhang, J. Vala, S. Sastry, and K. B. Whaley, Phys. Rev.
A {\bf 67}, 042313 (2003).

\bibitem{Fan}  H. Fan, V. Roychowdhury, and T. Szkopek, quant-ph/0410001.

\bibitem{Ye}  M.-Y. Ye, Y.-S. Zhang, and G.-C. Guo, quant-ph/0407108 v1.

\bibitem{Hami}  G. Vidal, K. Hammerer, and J. I. Cirac, Phys. Rev. Lett. 
{\bf 88}, 237902 (2002); H. L. Haselgrove, M. A. Nielsen, and T. J. Osborne,
Phys. Rev. A {\bf 68}, 042303 (2003); M. J. Bremner, D. Bacon, and M. A.
Nielsen, quant-ph/0405115.

\bibitem{Zeier}  R. Zeier, M. Grassl, and T. Beth, Phys. Rev. A {\bf 70},
032319 (2004).

Figure caption.

All gates represented by points in tetrahedron $ABCD$ can simulate any
two-qubit unitary operations with only 3 applications. Every point in
tetrahedron $ABCD$, $BCDE$ or $ABDE$ represents a perfect entangler. Here
the coordinates of points $O$, $A$, $B$, $C$, $D$, $E$ and $F$ are $\left(
0,0,0\right) $, $\left( \frac \pi 4,0,0\right) $, $\left( \frac \pi 4,\frac %
\pi 8,0\right) $, $\left( \frac \pi 4,\frac \pi 4,0\right) $, $\left( \frac %
\pi 8,\frac \pi 8,0\right) $, $\left( \frac \pi 8,\frac \pi 8,\frac \pi 8%
\right) $ and $\left( \frac \pi 4,\frac \pi 4,\frac \pi 4\right) $
respectively.
\end{references}
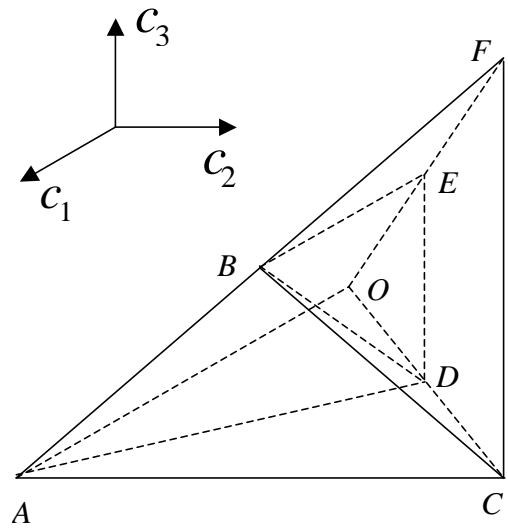
\end{document}